\documentclass[prd,nofootinbib,twocolumn]{revtex4}
\usepackage{graphicx}
\usepackage{bm}
\usepackage{float}
\usepackage{subfigure}
\usepackage{amsmath}
\usepackage{amsfonts}
\usepackage{color}
\usepackage{slashed}
\usepackage{fancyhdr}

\newcommand{\dd}{\text{d}}

\newcommand{\sbar}{\overline{\sigma}}

\begin{document}

\title{A Semi-Classical Schwinger-Keldysh Re-interpretation Of The 4D Majorana Fermion Mass Term}

\author{Yi-Zen Chu}
\affiliation{
	Department of Physics, National Central University, Chungli 32001, Taiwan;\footnote{Affiliation began 1 August 2017.} \\
	Department of Physics, University of Minnesota, Duluth, MN 55812, USA\footnote{Affiliation ended 22 May 2017.}
}
%\date{}

\begin{abstract}
	\noindent We offer a semi-classical re-interpretation of the 4D Majorana fermion mass term as an `influence action' in the Schwinger-Keldysh formulation of fermionic Quantum Field Theories.
\end{abstract}

\maketitle

\section{Motivation and Result}
\label{Section_Motivation}

Within the pedagogy of physics, the apparent necessity for anti-commuting Grassmann numbers when describing the 4-dimensional (4D) Majorana fermion mass term is sometimes explained by pointing out the associated Lagrangian density would otherwise vanish. The latter could, in turn, be traced to the antisymmetric nature of the charge conjugation matrix $C$. 

Now, if $L$ and $R$ are respectively two-component `left-handed' and `right-handed' spinors, the Dirac equations (written in the chiral basis) for a particle of mass $m > 0$, namely
\begin{align}
i\sbar^\mu \partial_\mu L 	&= m R \qquad \text{and} \\
i\sigma^\mu \partial_\mu R 	&= m L,
\end{align}
do admit a Lagrangian density involving only complex numbers, as encoded in the following equations:
\begin{align}
\label{DiracLagrangian}
\mathcal{L}_{\text{Dirac }m>0} 	&= \mathcal{L}_{L,0} + \mathcal{L}_{R,0} + \mathcal{L}_{\text{Dirac mass}} , \\ 
\label{LeftHandedWeySpinor}
\mathcal{L}_{L,0}[L,\partial L]	&\equiv L^\dagger i \sbar^\mu \partial_\mu L, \\ 
\mathcal{L}_{R,0}[R,\partial R]	&\equiv R^\dagger i \sigma^\mu \partial_\mu R, \\ 
\label{DiracMass}
\mathcal{L}_{\text{Dirac mass}} &\equiv - m \left( L^\dagger R + R^\dagger L \right) .
\end{align}
\footnote{{\it Notation:} In this paper we work exclusively in 4D Minkowski spacetime, where Cartesian coordinates shall be employed; Einstein summation covention is in force, with small Greek indices running over $0$ (time) and $1-3$ (space) while small Latin/English alphabets only run over the spatial ones $1-3$. The $\sigma^\mu \equiv (\mathbb{I}_{2 \times 2},\sigma^i)$, $\sbar^\mu \equiv (\mathbb{I}_{2 \times 2},-\sigma^i)$, $\mathbb{I}_{2 \times 2}$ is the $2 \times 2$ identity matrix, $\{ \sigma^i \vert i=1,2,3\}$ are the Pauli matrices obeying the algebra $\sigma^i \sigma^j = \delta^{ij} + i \epsilon^{ijk} \sigma^k$, and $\epsilon^{ijk}$ is the Levi-Civita symbol with $\epsilon^{123} \equiv 1$. The defining algebra of the charge conjugation matrix $C$ is: $C (\sbar^\mu)^* C^{-1} = \sigma^\mu$.}On the other hand, the Majorana equation 
\begin{align}
\label{MajoranaEquation}
i \sbar^\mu \partial_\mu L = m C L^* , 
\end{align}
where $^*$ denotes complex conjugation, appears at first sight to arise from 
\begin{align}
\mathcal{L}_{\text{Majorana }m>0} 	&= \mathcal{L}_{L,0} + \mathcal{L}_{\text{Majorana mass}} , \\  
\label{MajoranaMass_InOut}
\mathcal{L}_{\text{Majorana mass}} 	&\equiv -\frac{m}{2} \left( L^\dagger C L^* + L^T C^\dagger L \right) .
\end{align}
However, upon closer examination, one discovers $L^\dagger C L^* = L^T C^\dagger L = 0$ because
\begin{align}
\label{C_Antisymmetric}
C^T = -C ,
\end{align}
provided $L$ only involves complex numbers. Thus, it is at this point that $L$ is postulated to be built out of Grassmann numbers, so that $\mathcal{L}_{\text{Majorana mass}} \neq 0$.\footnote{See, for instance, problem 3.4 of Peskin and Schroeder \cite{Peskin:1995ev} and footnote 3 of \S II.1 of Zee \cite{Zee:2003mt}.}

While we are not disputing the Grassmannian nature of fermions upon their quantization, we were motivated -- while considering the appropriate semi-classical limits -- by the asymmetric treatment (i.e., why Grassmann numbers are introduced in one and not the other) of the Dirac mass term in eq. \eqref{DiracMass} versus that of the Majorana one in eq. \eqref{MajoranaMass_InOut}. In this note we offer an alternate action for the semi-classical Majorana theory of eq. \eqref{MajoranaEquation}:
\begin{align}
\label{MajoranaMass_InIn}
& S_{\text{``In-In" Majorana}} \nonumber\\
&\equiv \int_{t_\text{i}}^{t_\text{f}} \dd^4 x \Big( 
\mathcal{L}_{L,0}[L_\text{I},\partial L_\text{I}] 
- \mathcal{L}_{L,0}[L_\text{II},\partial L_\text{II}] \nonumber\\
&\qquad\qquad\qquad\qquad
+ \mathcal{L}_{\text{IF}}[L_{\text{I}},L_{\text{II}}] \Big) ,
\end{align}
with $\mathcal{L}_{L,0}$ already defined in eq. \eqref{LeftHandedWeySpinor} and the ``influence action" $\mathcal{L}_{\text{IF}}[L_{\text{I}},L_{\text{II}}]$ mixing the copy-``I" and copy-``II" of the same Majorana field $L$ given by the expression
\begin{align}
\label{MajoranaMass_L_InIn}
\mathcal{L}_{\text{IF}}[L_{\text{I}},L_{\text{II}}]
&\equiv -m \left( L_{\text{I}}^\dagger C L^*_{\text{II}} + L^T_{\text{II}} C^\dagger L_{\text{I}} \right) .
\end{align}
Moreover, the limits $t_{\text{i}}$ and $t_{\text{f}}$ in eq. \eqref{MajoranaMass_InIn} indicate the domain of the spacetime volume integral $\int \dd^4 x$ is bounded within some initial  $\Sigma[t_\text{i}]$ and final  $\Sigma[t_\text{f}]$ constant-time hypersurfaces. Finally, notice that $\mathcal{L}_{\text{IF}}$ is non-zero because of the presence of two distinct sets of fields $L_\text{I}$ and $L_\text{II}$; if $L_\text{I} = L_\text{II}$, the influence action $\mathcal{L}_{\text{IF}}$ would again be zero because of the antisymmetric nature of $C$, i.e., eq. \eqref{C_Antisymmetric}.

The form of the action principle in eq. \eqref{MajoranaMass_InIn} was inspired by the recent reformulation of Hamilton's principle in \cite{Galley:2012hx}, to incorporate retarded boundary conditions and dissipative dynamics.\footnote{See also \S V of Polonyi \cite{Polonyi:2012qw}. For a (small) sample of Schwinger-Keldysh applications in the classical limit that followed, see \cite{Grozdanov:2013dba} and \cite{Galley:2014wla}.} Such a ``doubling" of the fields to form an action is commonplace within the Schwinger-Keldysh/``In-In" formalism of Quantum Field Theory (QFT) for computing expectation values of quantum operators.

Before moving on, it should also be emphasized that the Majorana mass term in eq. \eqref{MajoranaMass_L_InIn}, because it couples the `I' and `II' fields, necessarily arises from the presence of an external agent or an environment. This does further distinguish our formulation here from the usual `in-out' Majorana mass term, which can exist within a closed quantum system. Furthermore, we reiterate that, to use eq. \eqref{MajoranaMass_InIn} to describe quantum many-body physics, one would still be imposing anti-commutation relations between $L$ and its conjugate momenta $\partial \mathcal{L}_{L,0}/\partial (\partial_t L)$; so that Majorana fermions will -- within the fully quantum domain -- obey Fermi-Dirac statistics/the Pauli exclusion principle, just like their Dirac fermion counterparts. We are merely suggesting a different semi-classical route Majorana fermions may be arrived at.

\section{Principle of Stationary Action: Doubled Fields For Majorana}
\label{Section_SchwingerKeldyshMajorana}
{\bf Setup} \qquad To set up a variational principle using the action in eq. \eqref{MajoranaMass_InIn}, we adopt the following prescription that is a minor modification of that in \cite{Galley:2012hx} due to the first-order nature of the fermionic system of partial differential equations (PDEs) at hand.\footnote{The justification of this procedure likely comes from taking the semi-classical limit, stationary-phase approximation, of the associated Schwinger-Keldysh path integral for Majorana fermions. Here, we shall focus solely on verifying that eq. \eqref{MajoranaEquation} is recovered from eq. \eqref{MajoranaMass_InIn}. }

{\it Perturbations} \qquad We begin by perturbing the fields; carrying out the replacements
\begin{align}
L_\text{I} 	&\to L_\text{I} + \delta L_\text{I}, \\
L_\text{II} &\to L_\text{II} + \delta L_\text{II} ;
\end{align}
and demanding that the first-order variation of $S_{\text{``In-in" Majorana}}$ -- the terms linear in the perturbations $\delta L_\text{I}$ and $\delta L_\text{II}$ -- be zero. 

{\it Boundary conditions} \qquad To proceed, we shall further assume that the field profiles have been specified on some initial constant-time hypersurface, which we denote as $\Sigma[t_\text{i}]$ and parametrize with coordinates $\{\vec{y}'\}$. In other words:
\begin{align}
\label{BoundaryConditions}
L_\text{I}[t_\text{i},\vec{y}'] 		&\text{ and } L_\text{II}[t_\text{i},\vec{y}'] \text{ specified}, \nonumber\\
\delta L_\text{I}[t_\text{i},\vec{y}'] 	&= \delta L_\text{II}[t_\text{i},\vec{y}'] = 0 . 
\end{align}
On the final constant-time hypersurface $\Sigma[t_\text{f}]$, again parametrized by $\{ \vec{y}' \}$, we will not demand that the ``I" and ``II" fields have been fixed, but that they coincide there:
\begin{align}
\label{JunctionConditions}
L_\text{I}[t_\text{f},,\vec{y}'] 		&= L_\text{II}[t_\text{f},,\vec{y}'] \qquad \text{and} \qquad \nonumber\\
\delta L_\text{I}[t_\text{f},\vec{y}'] 	&= \delta L_\text{II}[t_\text{f},\vec{y}'] .
\end{align}
{\it Physical limit} \qquad The last step in the procedure is to set the ``I" and ``II" fields equal in the ensuing equations-of-motion. (This is dubbed the ``physical limit" in \cite{Galley:2012hx, Galley:2014wla}.)

We note in passing that the Majorana action in eq. \eqref{MajoranaMass_InIn} is antisymmetric under the interchange of the labels $\text{I} \leftrightarrow \text{II}$; this antisymmetry is imposed on actions occurring in the Schwinger-Keldysh formulation of QFT.

{\bf Calculation} \qquad It is technically convenient to suppose we can find a unit norm future-directed timelike vector $u^\mu$ that is orthogonal to the initial/final time hypersurfaces $\Sigma[t_{\text{i},\text{f}}]$, such that the induced geometries on the latter are $h_{ij}[t_{\text{i},\text{f}},\vec{y}]$. For, carrying out the first order variation of eq. \eqref{MajoranaMass_InIn},
{\allowdisplaybreaks\begin{align}
\label{FirstOrderVariation}
0 &= \delta S_{\text{``In-In" Majorana}} \nonumber \\
&= \int_{t_\text{i}}^{t_\text{f}} \dd^4 x \delta L^*_\text{I A} \left( i (\sbar \cdot \partial)_{\text{AB}} L_\text{I B} - m  C_{\text{AB}} L^*_{\text{II B}} \right) \nonumber \\
&+ \int_{t_\text{i}}^{t_\text{f}} \dd^4 x \delta L_\text{I A} \left( -i (\sbar \cdot \partial)_{\text{BA}} L_\text{I B}^* + m  C_{\text{BA}}^* L_{\text{II B}} \right) \nonumber \\
&- \int_{t_\text{i}}^{t_\text{f}} \dd^4 x \delta L_\text{II A} \left( -i (\sbar \cdot \partial)_{\text{BA}} L_\text{II B}^* + m  C_{\text{BA}}^* L_{\text{I B}} \right) \nonumber \\
&- \int_{t_\text{i}}^{t_\text{f}} \dd^4 x \delta L_\text{II A}^* \left( i (\sbar \cdot \partial)_{\text{AB}} L_\text{II B} - m  C_{\text{AB}} L_{\text{I B}}^* \right) \nonumber \\
&+ \text{Boundary Terms} ;
\end{align}}
where the capital Latin/English alphabets run over the 2 spinor components; eq. \eqref{C_Antisymmetric} was employed to manipulate some of the terms; while the ``Boundary Terms" (BT$_\text{M}$) are
{\allowdisplaybreaks\begin{align}
\label{FirstOrderVariation_BC}
\text{BT}_{\text{M}} &\equiv \left(\int_{\Sigma[t'=t_\text{f}]} - \int_{\Sigma[t'=t_\text{i}]}\right) \dd^3\vec{y}' \sqrt{h'} \\
&\times \left( L^\dagger_{\text{I}} i u_\mu \sbar^\mu \delta L_{\text{I}} - L^\dagger_{\text{II}} i u_\mu \sbar^\mu \delta L_{\text{II}} \right) , \nonumber 
\end{align}}
with $h' \equiv \det h_{ij}[t',\vec{y}']$.

The boundary conditions in eq. \eqref{BoundaryConditions} set to zero the $t'=t_\text{i}$ terms of eq. \eqref{FirstOrderVariation_BC}; whereas the $t'=t_\text{f}$ ones are zero by eq. \eqref{JunctionConditions}. At this juncture, the principle of stationary action has lead us to deduce from equations \eqref{MajoranaMass_InIn} and \eqref{FirstOrderVariation} the pair of first-order PDEs:
\begin{align}
i \sbar^\mu \partial_\mu L_{\text{I}} 	&= m C L_\text{II}^* \qquad \text{and} \\
i \sbar^\mu \partial_\mu L_{\text{II}} 	&= m C L_\text{I}^*  .
\end{align}
Upon imposing the `physical limit' \cite{Galley:2012hx,Galley:2014wla}, $L_\text{I} = L_\text{II}$, we recover the desired Majorana theory of eq. \eqref{MajoranaEquation}.

{\bf Remarks} \qquad The difference between first order systems such as the Dirac or Majorana fermion and second order ones commonly encountered in bosonic systems is that, in the latter, two conditions are needed -- usually, either boundary values or the initial field configuration and its velocity are specified -- for the solution to the relevant differential equations be determined uniquely. Whereas, for the former, only one is required: normally, it is the initial wave function ($L$ for the Majorana fermion) that is given.

Specifically, the usual ``In-Out" variational principle for the Dirac fermion in eq. \eqref{DiracLagrangian} admits the following first order perturbation:
{\allowdisplaybreaks\begin{align}
&\delta \int_{t_\text{i}}^{t_\text{f}} \dd^4 x \mathcal{L}_{\text{Dirac $m>0$}} \nonumber\\
&= \int_{t_\text{i}}^{t_\text{f}} \dd^4 x \delta L^\dagger \left( i (\sbar \cdot \partial) L - m R \right) \nonumber\\
&+ \int_{t_\text{i}}^{t_\text{f}} \dd^4 x \left( -i \partial_\mu L^\dagger \sbar^\mu - m R^\dagger \right) \delta L \nonumber\\
&+ \int_{t_\text{i}}^{t_\text{f}} \dd^4 x \delta R^\dagger \left( i (\sigma \cdot \partial) R - m L \right) \nonumber\\
&+ \int_{t_\text{i}}^{t_\text{f}} \dd^4 x \left( -i \partial_\mu R^\dagger \sigma^\mu - m L^\dagger \right) \delta R \nonumber\\
&+ (\text{Boundary Terms})'
\end{align}}
where the ``$(\text{Boundary Terms})'$" (BT$_\text{D}$) reads
{\allowdisplaybreaks\begin{align}
\label{FirstOrderVariationPrime_BC}
\text{BT}_{\text{D}} &\equiv \left(\int_{\Sigma[t'=t_\text{f}]} - \int_{\Sigma[t'=t_\text{i}]}\right) \dd^3\vec{y}' \sqrt{h'} \\
&\times \left( L^\dagger i u_\mu \sbar^\mu \delta L + R^\dagger i u_\mu \sigma^\mu \delta R \right) . \nonumber 
\end{align}}
To define a principle of stationary action for the Dirac fermion, one would further impose the boundary conditions that the wave function be fixed on the initial and final constant-time hypersurfaces $\Sigma[t_{\text{i},\text{f}}]$,
\begin{align}
\label{FirstOrderVariationPrime_LR_BC}
\delta L[t_\text{i},\vec{y}'] = \delta R[t_\text{i},\vec{y}'] 
= \delta L[t_\text{f},\vec{y}'] = \delta R[t_\text{f},\vec{y}'] = 0 .
\end{align}
The absence of any derivatives in the boundary terms of the Majorana eq. \eqref{FirstOrderVariation_BC} and its Dirac counterpart of eq. \eqref{FirstOrderVariationPrime_BC} is directly linked to the first-order nature of both theories. Note, in particular, that once $L[t_\text{i},\vec{y}']$ and $R[t_\text{i},\vec{y}']$ are given, there is no freedom is choosing $L[t_\text{f},\vec{y}']$ and $R[t_\text{f},\vec{y}']$ -- i.e., $\delta L[t_\text{f},\vec{y}'] = \delta R[t_\text{f},\vec{y}'] = 0$ automatically -- and hence the two rightmost equalities of eq. \eqref{FirstOrderVariationPrime_LR_BC} are actually redundant.

\section{Discussions and Future Directions}
\label{Section_Discussion}
In this paper we have offered a novel ``In-In"-QFT inspired \cite{Galley:2012hx} starting point for a 2-component fermion with a Majorana mass term in 4D flat spacetime. At the level of the one-particle sector -- i.e., the semi-classical equations-of-motion of eq. \eqref{MajoranaEquation} -- the action formulation in eq. \eqref{MajoranaMass_InOut} in terms of a Grassmannian $L$ versus that of eq. \eqref{MajoranaMass_InIn} in terms of a complex-valued influence-action really yield no physical distinctions. However, we hope such a re-phrasing of its action principle could would spur further explorations. For instance, whether equations \eqref{MajoranaMass_InOut} and \eqref{MajoranaMass_InIn} do in fact lead to different physical observables in the full quantum theory of Majorana fermions. How does the Majorana mass term in eq. \eqref{MajoranaMass_InIn} arise from interactions with an external environment or external agent? We also note, within the ``In-In" (aka Closed Time Path) formalism, that the ``I" fields are usually interpreted as ones propagating forward in time whereas the ``II" fields backward in time. To this end, the study of both discrete and continuous symmetries of the Majorana mass term of eq. \eqref{MajoranaMass_L_InIn} could perhaps enrich our understanding of its physical content. We shall also continue to seek out possible physical applications.

\section{Acknowledgments}

I wish to thank Nishant Agarwal, Saso Grozdanov, Archana Kamal and especially Daniel Schubring for useful comments and discussions. I also thank the anonymous referee for constructive criticisms. A portion of the work presented here was done while I was employed at the University of Minnesota Duluth.

\appendix

\end{document}